\documentstyle[prl,aps]{revtex}
\begin{document}
\draft
\twocolumn
\title{\bf Polarization of Instantons and Gravity}
\author{M.~Yu.~Kuchiev\cite{byline}}
\address{School of Physics, University of New South Wales,
Sydney, 2052, Australia}
\date{\today}
\maketitle
\begin{abstract}
Gravity can arise in a conventional non-Abelian gauge theory in which
a specific phenomenon takes place. Suppose there is a condensation of
polarized instantons and antiinstantons in the vacuum state. Then the
excitations of the gauge field in the classical approximation are described
through the variables of Riemann geometry satisfying the Einstein equations
of general relativity. There are no dimensional coupling constants in the
theory.
\end{abstract}

In this work I want to demonstrate that conventional gauge
theory can provide the basis for gravity.
The theory under consideration is the
usual non-Abelian gauge field theory \cite{YM} dealing
exclusively with  particles of spin 0, 1/2 and 1. There
are no gravitons on the basic level. Space-time is fundamentally
flat. There are no dimensional coupling constants in the theory.

Obviously there must appear something nontrivial in the usual gauge
theory which yields the gravitational effect. This is
a specific phenomenon in which instantons \cite{Bel} play a
crucial role. Generally speaking an instanton  may be oriented arbitrarily.
Let us assume that the vacuum state possesses a condensate of instantons
with
a specific property: the instantons belonging to this condensate have a
preferred orientation. We shall call them ``polarized'' instantons. More
precisely we suppose the existence of two condensates. One condensate
consists of polarized instantons belonging to some $su(2)$ sub-algebra of
the
gauge algebra. The other is a condensate of polarized antiinstantons
belonging
to some other $su(2)$ sub-algebra of the gauge algebra. We
shall call this property the Instanton-Antiinstanton Polarization (IAP).

Suppose that there
is a non-Abelian gauge theory in which there is IAP in the vacuum
state. Then this theory is shown to describe the effects of gravity.
Gravity arises due to a smooth variation of the orientation of the
condensates
of instantons and antiinstantons in space-time. The
main quantities which describe gravity - the metric, Christoffel
symbols, Riemann tensor - are functions of the gauge field. The
Einstein equations of general relativity are shown to arise as a
classical approximation for this gauge theory. These
statements are valid on the classical level only. The basic
quantum theory is  the theory of the gauge field in flat space-time.
The possibility of constructing a gauge theory with the IAP
is not discussed in this paper, - it needs special
consideration.

The mathematical resemblance between the variables in general
relativity and in gauge theories is well known, see [3]. In [4]
there was developed the approach which in some cases helps to
clarify this resemblance. Certainly there are distinctions.
 The most important for our consideration are
the following:\\ i. General relativity is invariant under the
local Lorentz transformations. The Lorentz group is non-compact.
In contrast to that the successful gauge theories are based upon a
compact group of gauge transformations.\\ ii. The Lagrangian for
the gravitational field in general relativity is linear with
respect to the Riemann tensor (defined in accordance with \cite{LL})
\begin{equation} \label{lg}
{\cal L}_{GR} = - \frac{1}{16 \pi k}\sqrt{-g}~ g^{\lambda \mu}g^{
\rho \nu}R_{\lambda \rho \mu \nu}~,
\end{equation}
while the Lagrangian for the gauge field is quadratic with respect
to $F^{ij}_{\mu \nu}$,
\begin{equation} \label{lym}
{\cal L}_{YM} = - \frac{1}{4g^2} F^{ij}_{\mu \nu} F^{ij}_{\mu
\nu}~.
\end{equation}
The obvious resemblance between the Riemann tensor
and the strength of a gauge field does not
manifest itself on the level of the Lagrangians.This distinction
makes the  classical equations of motion
in the two theories to be quite different.
In gravity theories
with higher covariant derivatives, see  \cite{hc},
the Lagrangian is more sophisticated then in Eq.(\ref{lg}),
but still the linear in $R_{\lambda \rho \mu \nu}$ term plays a role,
preventing the gravity Lagrangian to be equal to the Yang-Mills
Lagrangian.\\
iii. The basic quantity in general relativity is the metric, the
Christoffel symbols $\Gamma^{\lambda}_{\mu \nu}$ being derived
from it. On the other hand in gauge theories the vector potential
$A^{ij}_{\mu}$ plays the major role. Again, the resemblance between
the Christoffel symbols and the vector potential is not manifested in the
dynamical properties of the two theories.

We will see that IAP is a remedy which sweeps away
all these distinctions.
The first question to be addressed is the gauge
group.
The local Lorentz group is characterized by six parameters. If
a gauge theory is believed to describe the effects of gravity
then the number of parameters of the gauge
group is to be six (or more). The simplest compact group with six
parameters is SO(4). Therefore let us consider the usual SO(4)
gauge theory with the conventional Lagrangian which includes the
term (\ref{lym}) describing the non-Abelian gauge field. Gauge theories
may differ in the number of generations of the scalars and
fermions, their masses and coupling constants which results in a
variety of properties of the theories. In this work we assume that among
these
theories
there exists a gauge theory with the necessary property - IAP.

The considered gauge algebra $so(\ref{sins})$ is the direct sum of two
$su(\ref{lym}), so(\ref{sins}) = su(\ref{lym}) + su(\ref{lym})$. We can
choose
the generators for one $su(\ref{lym})$ to be $(- 1/2)\eta_{aij}$ and the
generators for the other to be $(-1/2) \bar \eta_{aij}$, and refer to these
algebras as $su(\ref{lym})
\eta$ and $su(\ref{lym}) \bar \eta$. Here $\eta_{aij}, \bar \eta_{aij}$
are the 't Hooft symbols, $a =1,2,3, ij = 1, \cdots, 4$. The
strength of the gauge field in this notation is
\begin{equation} \label{F}
F^{ij}_{\mu \nu}=-(1/2)(F^a_{\mu \nu} \eta_{aij} + \bar F^a_{\mu
\nu} \bar \eta_{aij})~,
\end{equation}
where $F^a_{\mu \nu}$ belongs to $su(\ref{lym}) \eta$ and $\bar F^a_{\mu
\nu}$ belongs to $su(\ref{lym}) \bar \eta$.

We assume that there is a condensate of polarized instantons belonging to
$su(\ref{lym}) \bar \eta$ and a condensate of polarized antiinstantons
belonging to $su(\ref{lym}) \eta$. Now let us consider the
interaction of the sufficiently weak and slowly varying gauge
field with this polarized state. From now on the Euclidean
formulation is used. The interaction of one
instanton with a smoothly varying weak field was considered
in Ref.\cite{CDG}, see also \cite{VZNS}. For the case of the $SU(2)$ gauge
group the effective action describing this interaction is
\begin{equation} \label{sins}
S_{ins}=(2 \pi^2 \rho^2/g^2)~ \bar \eta_{a \mu \nu} \bar F^b_{\mu
\nu} \bar C_{ab}~,
\end{equation}
where $\rho$ is the radius of instanton, and $\bar C_{ab} \in SO(3)$ is a
matrix describing the orientation of the instanton. The bar over
$\bar F^b_{\mu \nu}$ reminds us that the instanton under
consideration belongs to $su(\ref{lym}) \bar \eta$. Eq.(\ref{sins}) remains
valid for the antiinstanton as well if we substitute $\bar \eta_{a \mu
\nu} \rightarrow \eta_{a \mu \nu}, \bar F^b_{\mu \nu} \rightarrow
F^b_{\mu \nu}$, and $\bar C_{ab} \rightarrow C_{ab}$, where $C_{ab
} \in SO(3)$ is a matrix describing the orientation of the antiinstanton.

Now let us apply this result to the vacuum with IAP. If we suppose the
dilute gas approximation for the polarized instantons and antiinstantons to
be
valid then from Eq.(\ref{sins}) with
the help of Eq.(\ref{F}) we deduce that the interaction of the slowly
varying weak field with this vacuum is described by the Lagrangian
\begin{equation} \label{lpol}
{\cal L} = - (\eta_{a \mu
\nu} \eta_{bij} M_{ab} + \bar \eta_{a \mu \nu} \bar \eta_{bij} \bar M_{ab})
F^{ij}_{\mu \nu}~.
\end{equation}
Here  $M_{ab}, \bar M_{ab}$ are defined as
\begin{eqnarray} \label{M}
M_{ab} = \pi^2~\langle ~ (1 /g^2) \rho^2 n(\rho,C) C_{ab}~\rangle~,
\\ \label{bM}
\bar M_{ab} =
\pi^2~\langle ~(1 /g^2) \rho^2 \bar n(\rho,\bar C) \bar C_{ab}~
\rangle~,
\end{eqnarray}
where $n(\rho,C)$ is the concentration of the $su(2)\bar \eta$ instantons
with
radius $\rho$ and orientation given by the matrix $C_{ab}$, and
$\bar n(\rho,\bar C)$ is the concentration of the $su(2)\eta$ antiinstantons
with orientation given by the matrix $\bar C_{ab}$. The brackets $\langle~
\rangle$ in (\ref{M}),(\ref{bM}) describe
the average over microscopic fluctuations of the gauge field.
The existence of IAP means that there are condensates of polarized
$su(2)\bar \eta$ instantons and $su(2)\eta$ antiinstantons
which give the nonzero contribution to the right-hand sides of (\ref{M}),
(\ref{bM})
\begin{eqnarray} \label{fc}
M_{ab} = (f/4)~C^{{\it cond}}_{ab}~,~~
\bar M_{ab} = (\bar f/4)~\bar C^{{\it cond}}_{ab}~,
\end{eqnarray}
where $C^{{\it cond}}_{ab} \in SO(3)$ is the matrix describing the
orientation
of the condensate of polarized instantons and $\bar C^{{\it cond}}_{ab} \in
SO(3)$ is the matrix describing the orientation of the condensate of
polarized
antiinstantons. The constants $f,\bar f$ characterize the intensity of these
condensates. We will suppose them to be equal, and hence $f = \bar f$.

It is useful to present the matrixes
$C^{{\it cond}}_{ab}, \bar C^{{\it cond}}_{ab}$ with
the help
of a  matrix $h^{ij} \in SO(4)$ which satisfies the
equations
\begin{eqnarray} \label{hh}
h^{ik}h^{jl} \eta_{akl} = C^{{\it cond}}_{ab} \eta_{bij}~,~~
h^{ik}h^{jl} \bar \eta_{akl} = \bar C^{{\it cond}}_{ab} \bar \eta_{bij}~,
\end{eqnarray}
and describes the orientation of the condensate of instantons
and antiinstantons.

Substituting Eqs.(\ref{fc}),(\ref{hh}) into Eq.(\ref{lpol}) and using the
identity \cite{th}
\begin{equation} \label{eta}
\eta_{a \mu \nu} \eta_{aij}+ \bar \eta_{a \mu \nu} \bar \eta_{aij}
= 2(\delta_{i \mu} \delta_{j \nu}- \delta_{j \mu} \delta_{i \nu})~,
\end{equation}
one finds that the Lagrangian (\ref{lpol}) may be written as
\begin{equation} \label{dlym}
\Delta {\cal L}_{YM} = - f h^{i \mu} h^{j \nu}
F^{ij}_{\mu \nu}~.
\end{equation}
Remember that the Latin letters $i,j$ label the indexes of variables in the
isotopic
space while the Greek letters $\mu,\nu$ label the indexes in the coordinate
space.
The symbols $\eta_{aij}$ play a role of the generators of the gauge
transformations, see Eq.(\ref{F}), while the symbols $\eta_{a\mu\nu}$
describe
the orientation of instantons in  the coordinate space, see Eq.(\ref{sins}).
Eq.(\ref{eta}) gives a match between the indexes of isotopic and coordinate
spaces. It makes it useful to consider the
matrix $h^{i\mu}$ in Eq.(\ref{dlym}) with one Latin index and one Greek one.
  This matrix plays a role of the order parameter of the problem.

{}From Eq.(\ref{dlym}) one deduces that there appears the corresponding term
in
the action:
\begin{equation} \label{dels}
\Delta S= - f \int h^{i \mu}h^{j \nu} F^{ij}_{\mu \nu}
\det h~d^4x~.
\end{equation}
This action is invariant under two types of transformations. First,
it preserves gauge symmetry. Gauge transformations have a form
\begin{eqnarray} \label{Fgaug}
F'^{ij}_{\mu \nu} =U^{ik}(x) U^{jl}(x) F^{kl}_{\mu \nu}~,~~
h'^{i\mu} = U^{ki}(x) h^{k\mu}~.
\end{eqnarray}
Second, it  is invariant under the transformations of the coordinates
$x^{\mu} \rightarrow x'^{\mu}$
\begin{eqnarray} \label{Fxx}
F'^{ij}_{\mu \nu} =\frac{\partial x^{\rho}} {\partial x'^{\mu}}
\frac{\partial x^{\sigma}}{\partial x'^{\nu}}
F^{ij}_{\rho \sigma}~,~~
h'^{i\mu} =  \frac{\partial x'^{\mu}}{\partial
x^{\nu}}h^{i\nu}~.
\end{eqnarray}
The factor $\det h (\equiv \det h^{i}_{\mu}$, where $h^{i}_{\mu}$
denotes the matrix inverse to $h^{i \mu}:~ h^{i \mu}h^{j}_{\mu}=
\delta_{ij}$) in Eq.(\ref{dels}) compensates for the variation of the
volume $d^4x$  due to the identity
$\det h = \det{[\partial x^{\mu}/ \partial x'^{\nu}]}$,
which follows from Eq.(\ref{Fxx}).

Up to now we assumed the condensate to be homogeneous and
the matrix $h^{i\mu}$ was considered as a global orthogonal matrix,
$h^{i\mu} \in SO(4)$.  This  is true for some particular  coordinate frame
and
particular  gauge condition, see
(\ref{Fgaug}),(\ref{Fxx}).
Similar consideration may be fulfilled for the case
of the non-homogeneous condensate as well. In this case the order parameter
may be shown to be described by the matrix $h^{i\mu}(x)$ as well,
but there are two important distinctions.
First, the  matrix  $h^{i\mu}(x)$   varies in  space.
For non-homogeneous condensate this variation is nontrivial,
i.e.  it can not be eliminated with the help of the
transformations (\ref{Fgaug}),(\ref{Fxx}).
Second, $h^{i\mu}(x)$
is an arbitrary matrix  having 16 independent parameters.

Consider the interaction of a weak and smooth gauge field with the
non-homogeneous condensate which also varies smoothly in space.
Then it may be shown that this interaction
is described  by the same action
(\ref{dels}) assuming that $h^{i\mu}=h^{i\mu}(x)$ is a function of $x$.
A simple  argument in favour of this result is a fact that locally, at any
point $x_0$,  the order parameter $h^{i\mu}(x_0)$
may be transformed with the help of the coordinate transformation
(\ref{Fxx}) to be an orthogonal  matrix  $h^{i\mu}(x_0) \in SO(4)$.
Then the expression (\ref{dlym}) for the Lagrangian is valid in the vicinity
of $x_0$.  The point $x_0$ is arbitrary,  therefore we can integrate the
Lagrangian evaluating  the action (\ref{dels}).

The action (\ref{dels}) depends on the vector potential
$A^{ij}_{\mu}(x)$ and the matrix $h^{i \mu}(x)$.
$A^{ij}_{\mu}(x)$ is a slowly
varying vector potential
having the
trivial topological structure (at least on the microscopic level).
In contrast to that $h^{i \mu}(x)$ describes the orientation of
polarized instantons and antiinstantons which are
the degrees of freedom of the field with nontrivial
topological structure. This allows us to consider
$A^{ij}_{\mu}(x)$ and $h^{i \mu}(x)$ as separate variables.
Note that for the considered weak field
the term quadratic  in $F^{ij}_{\mu\nu}$  given by
the Lagrangian (\ref{lym}) is
much smaller compared to the linear term (\ref{dels}).  Therefore
we can neglect the action coming from Lagrangian (\ref{lym}).

The weak and smooth nature of
the field permits one to use the classical approximation:
\begin{eqnarray} \label{dsda}
\delta( \Delta S)/ \delta A^{ij}_{\mu} &=& 0~,
\\ \label{dsdh}
\delta( \Delta S)/ \delta h^{i \mu} &=& 0~.
\end{eqnarray}
Eq.(\ref{dsda}) gives the relation
between $h^{i \mu}$ and $A^{ij}_{\mu}$
\begin{equation} \label{dhhh}
\nabla_{\mu}((h^{i \mu }h^{j \nu }-h^{j \mu}h^{i \nu}) \det h)=0~.
\end{equation}
Here $\nabla_\mu$ is the covariant derivative in the gauge field
$(\nabla_{\mu})^{ij}= \delta_{ij} \partial_{\mu}+A^{ij}_{\mu}$. In order to
present Eq.(\ref {dhhh}) in a more convenient form
let us define three quantities, $g_{\mu \nu},
\Gamma^{\lambda}_{\mu \nu},$ and $R^{\lambda}_{\rho \mu \nu}$:
\begin{eqnarray} \label{ghh}
g_{\mu \nu} &=& h^{i}_{\mu} h^{i}_{\nu}~,
\\ \label{gam}
 \Gamma^{\lambda}_{\mu \nu} &=& h^{i \lambda}h^{j}_{\mu}A^{ij}_
{\nu} + h^{i \lambda} \partial_{\nu} h^{i}_{\mu}~,
\\ \label{RF}
R^{\lambda}_{\rho \mu \nu} &=& h^{i \lambda} h^{j}_{\rho}F^{ij}_
{\mu \nu}~.
\end{eqnarray}
Remember that the space-time under consideration is basically flat and
Eqs.(\ref{ghh}),(\ref{gam}),(\ref{RF}) define
the left-hand sides.

{}From (\ref{ghh}),(\ref{gam}) we find
that Eq.(\ref{dhhh}) may be presented in the  form:
$g_{\lambda \sigma} \Gamma^{\lambda}_{\mu \nu} = (1/2)
(\partial_{\mu}g_{\sigma \nu} + \partial_{\nu}g_{\sigma \mu} -
\partial_{\sigma}g_{\mu \nu})$.
demonstrating that we may consider $g_{\mu \nu}$ as a
 metric and $\Gamma^{\lambda}_{\mu \nu}$ as a Christoffel
symbol. Moreover, one finds that the quantity $R^{\lambda}_{\rho
\mu \nu}$ introduced in Eq.(\ref{RF}) turns out to be equal to the
Riemann tensor:
$R^{\lambda}_{\rho \mu \nu} = \partial_{\mu} \Gamma^{
\lambda}_{\rho \nu} - \partial_{\nu} \Gamma^{\lambda}_{\rho \mu}
+ \Gamma^{\lambda}_{\sigma \mu} \Gamma^{\sigma}_{\rho \nu} -
\Gamma^{\lambda}_{\sigma \nu} \Gamma^{\sigma}_{\rho \mu}$.
Considering now the second classical equation (\ref{dsdh}) one verifies
with the help of Eqs.(\ref{ghh}),(\ref{RF}) that it results in the Einstein
equations of general relativity in the absence of matter:
$R_{\mu \nu} - (1/2) g_{\mu \nu} R = 0$.

We come to the important conclusion.
If IAP takes place in the SO(\ref{sins}) gauge theory then the
classical approximation for this gauge theory may be described via the
variables of Riemann geometry for which the Einstein equations are valid.
These equations imply in particular that there exist gravitational waves.
That is a  pleasant surprise since the initial gauge theory possesses
no graviton
on the basic level. Graviton appears due to excitation of the condensate
of instantons and antiinstantons.

Consider the action (\ref{dels}) when the classical Eq.
(\ref{dsda}) is valid. It is clear from (\ref{ghh}),(\ref{gam}),(\ref{RF})
that it is identical to the Lagrangian of general relativity (\ref{lg}).
The gravitational constant turns out to be
\begin{equation} \label{newt}
k^{-1} = 16 \pi f~.
\end{equation}
This relation shows  that a radius and separation
of instantons which give the contribution to the constant $f$, see
Eqs.(\ref{M}),(\ref{fc}), are comparable to the Plank radius.

We see how the distinctions (i), (ii), (iii)
discussed at the beginning of the paper are eliminated by IAP. The
major problem is (i), the compactness of the group of local transformations.
IAP solves it in a peculiar manner. The variables of the gauge theory are
expressed, see (\ref{ghh}),(\ref{gam}),(\ref{RF}), via the variables of
Riemann geometry. This geometry deals with the
invariance with respect to transformations of a local reference frame. These
transformations may be continued from Euclidean to Minkowsky space.

I thank my colleagues at A.F.Ioffe Institute for a
stimulating scientific atmosphere. I thank
V.~V.~Flambaum, C.~J.~Hamer, and O.~P.~Sushkov for
helpful discussions. I appreciate the financial support from DITAC which
made
it possible to finish this work at School of Physics of UNSW, and the
hospitality of their staff is acknowledged. The kind help of L.S.Kuchieva on
all the stages of this work was vital.

{\it(The paper is accepted for publication in Europhysics Letters,
submitted in the final form 19 October, 1994.)}

\end{document}